# Molecular beam epitaxy of GaN/AlGaN quantum wells on bulk GaN substrate in the step-flow or step meandering regime: influence on indirect exciton diffusion

B. Damilano,[1*] R. Aristégui,[2] H. Teisseyre,[1,3,4] S. Vézian,[1] V. Guigoz,[1] A. Courville,[1] I. Florea,[1] P. Vennéguès,[1] M. Bockowski,[4] T. Guillet,[2] M. Vladimirova[2]

[1]Université Côte d'Azur, CNRS, CRHEA, Valbonne, France
[2]Laboratoire Charles Coulomb (L2C), University of Montpellier, CNRS, Montpellier, France
[3]Institute of Physics, Polish Academy of Sciences, Warsaw, Poland
[4]Institute of High-Pressure Physics, Polish Academy of Sciences, Warsaw, Poland

[*]Corresponding author, bd@crhea.cnrs.fr

## Abstract

GaN/Al$_x$Ga$_{1-x}$N quantum wells were grown by molecular beam epitaxy on high quality bulk (0001) GaN substrates. The quantum well thickness was set in the 6-8 nm range to favor the photoluminescence emission of indirect excitons. Indeed, such excitons are known to be spatially indirect, due to the presence of the internal electric field which spatially separates the electron and hole wave functions. The growth conditions were optimized in view of minimizing the photoluminescence peak broadening. In particular, the impact of growth temperature (up to 900°C) on the surface morphology, structural and photoluminescence properties was studied. The diffusion of indirect excitons on the scale of tens of microns was measured with a micro-photoluminescence setup equipped with a spatially resolved detection. A dedicated model and its analysis allow us to extract from these measurements the exciton diffusion constant and to conclude on the optimum growth conditions for the GaN/Al$_x$Ga$_{1-x}$N quantum well structures suited for studies of quantum collective effects in indirect exciton liquids.

# 1 Introduction

GaN/AlGaN quantum wells (QWs) are commonly used for a variety of devices or applications such as UV laser diodes,[1,2,3,4] UV light emitting diodes,[5,6] room temperature polariton lasers,[7] micro-disk or micro-ring resonators,[8,9] ridge waveguides,[10,11,12] intersubband optoelectronic in the infrared and the THz ranges,[13] thermal emitters coupled to photonic crystal.[14]

Most of the papers regarding GaN/AlGaN QWs were based on structures grown by heteroepitaxy on sapphire or silicon substrates using metalorganic vapor phase epitaxy (MOVPE)[15,16] or molecular beam epitaxy (MBE).[17,18] These structures are affected by a large density of threading dislocations typically in the $10^8$-$10^{10}$ cm$^{-2}$ range. The negative impact of these defects was highlighted on different properties such as the internal quantum efficiency of GaN/AlGaN QWs.[19] For example, it was shown that the room temperature photoluminescence (PL) intensity of a 8 mono-layers-thick GaN / Al$_{0.1}$Ga$_{0.9}$N QW was 60 times larger when grown on bulk GaN substrate rather than on sapphire substrate.[20] Many fundamental properties of GaN/AlGaN QWs were elucidated despite the presence of a high defect density: in particular the major influence of the quantum confined Stark effect on the PL properties.[21,22,23,24,25,26,27,28,29,30,31,32,33]

However, a large dislocation density is a major issue for applications requiring long excitonic radiative lifetimes ($t_{rad}$) of the GaN/AlGaN QWs, since the ratio giving the internal quantum efficiency IQE=$(1+t_{rad}/t_{nr})^{-1}$ becomes very small for a fixed non-radiative lifetime ($t_{nr}$). This situation occurs in wide GaN QWs (> 5 nm) that are requested to obtain dipolar or indirect excitons (IX).[34,35,36,37] Such excitons are promising for studies of quantum collective states that these excitons can form, including Bose-Einstein condensation, that has been reported for trapped GaAs/AlGaAs IXs.[38] However, such condensation has still not been demonstrated in the GaN/AlGaN system which holds the promise to grant access to condensates at higher temperatures and with higher exciton densities than in the GaAs/AlGaAs system.

Compared to nanostructures based on gallium arsenide (GaAs), the choice of gallium nitride is motivated by two factors. First, this material is expected to enable collective quantum effects at higher temperatures due to higher exciton binding energy (26 meV in the GaN massive material[39] in comparison to 4.2 meV for GaAs[40]) and smaller Bohr radius (3 nm in bulk GaN and 13 nm in GaAs).[41,42] Second, due to spontaneous and piezoelectric polarizations, GaN/AlGaN QWs are subject to strong build-in electric field, which pushes electrons and holes within the excitons towards opposite interfaces of the QW. Therefore, such excitons become spatially indirect even in the absence of any externally applied electric bias. This is an advantage compared to AlGaAs/GaAs QWs were application of an electric field in the growth direction is mandatory to observe IX optical emission,[43] which leads to the injection of spurious charge carriers, known to be detrimental for IX coherence properties.[44,45,46]

Nevertheless, important challenges must be faced to achieve IX condensation in GaN/AlGaN QWs, and, in particular, IX non-radiative emission rates need to be minimized. Using high quality GaN substrates is therefore a prerequisite to get efficient light emission from IXs in GaN/AlGaN QWs. This was previously illustrated by demonstrating room temperature transport of IXs for a GaN/AlGaN QW structure grown on a free-standing GaN substrate, while this transport was undetectable for the same structure grown on sapphire substrate.[35] Until now no specific study, investigating the growth parameters was realized to enhance the quality of GaN/AlGaN QWs for IX transport.

The purpose of this work is then to optimize and characterize wide GaN/AlGaN QWs whose properties are adapted to the propagation of IXs. Indeed, all sources of inhomogeneities, and non-radiative recombinations have to be avoided. This is all the more difficult since this should be obtained on a scale of several tenths of micrometers, the typical distance for IX diffusion. This is the reason why, in this work, we exclusively used low dislocation density ($<10^4$ cm$^{-2}$) GaN bulk substrates in order to increase, as far as possible, the quality of the epitaxial structures.

## 2 Experimental

The samples were grown by MBE on bulk (0001) ammonothermal GaN substrates (1x1 cm$^{-2}$ or 1 in. diameter) from Ammono with an offcut of 0.5° towards the m-plane.[47,48] The dislocation density of these substrates is estimated to be below $5 \times 10^4$ cm$^{-2}$. Two sample sets were grown using different substrate preparations before the MBE. For the first one, the substrates are de-oxidized using a HF-based solution. For the second one, the GaN substrates are overgrown with a 1.1 µm GaN layer grown by metal organic vapor phase epitaxy (MOVPE) at 1000°C. This second approach leads to smoother surfaces as will be shown in the following. The samples are mounted onto Indium-free 3-inch molyblock holders adapted to the shapes of the substrates. NH$_3$ is used as the nitrogen source. Ga and Al atoms are evaporated from a double filament effusion cell and a cold-lip effusion cell, respectively. The growth temperature is deduced from the sublimation rate of GaN under vacuum measured on calibration samples.[49]

The samples were initially heated at 780°C under an ammonia flow of 100 standard centimeter cubes per minute (sccm) for 10 minutes in order to remove residual surface contaminants such as C and O. A 600 nm GaN buffer layer was grown before a 100 nm-thick Al$_x$Ga$_{1-x}$N layer, a GaN QW and a 50 nm-thick Al$_x$Ga$_{1-x}$N top layer. For the first series of samples, the growth temperature was set at 743°C (sample A), 773°C (sample B), 796°C (sample C), and 818°C (sample D). For the second series of samples (with a MOVPE GaN buffer layer), the growth temperature was set at 780°C for sample A' and 900°C for sample B'. For this last sample, the surface was annealed under ammonia at 900°C for 20 min before the growth of the first Al$_x$Ga$_{1-x}$N layer. All the data regarding the samples can be found in Table 1.

Table 1. Growth conditions of the GaN/AlGaN layers, the Al composition ($x_{Al}$) and the quantum well thickness for the first set (A-D) and the second set (A' and B') of samples.

| Sample | Substrate preparation before MBE | Growth temperature (°C) | NH$_3$ flow (sccm) | $x_{Al}$ | QW thickness (nm) |
|---|---|---|---|---|---|
| A | Chemical | 743 | 100 | 0.082 | 7.8 |
| B | Chemical | 773 | 100 | 0.081 | 7.8 |
| C | Chemical | 796 | 100 | 0.084 | 7.8 |
| D | Chemical | 818 | 100 | 0.078 | 7.8 |
| A' | MOVPE GaN | 780 | 100 | 0.096 | 6.4 |
| B' | MOVPE GaN | 900 | 500 | 0.108 | 6.4 |

Atomic force microscopy in tapping mode was used to characterize the surface of the samples.

The high-resolution cathodoluminescence was performed with a MonoCL4 GATAN system equipped with a high-sensitivity photomultiplier mounted on a field emission gun scanning electron microscope (FEG-SEM; JEOL JSM700F). The electron beam current ranges typically from 1 to 4 nA and the voltage beam is fixed at 5 keV. Dispersed light is collected through a paraboloidal mirror and quantified as a whole using a CCD detector with a sensitivity range from 250 to 1050 nm. Acquisition time is fixed to 350 µs/pixel. All observations are done at room temperature.

Morphological and structural analyses of the considered samples were performed using a ThermoFisher Titan SPECTRA 200 transmission electron microscope (TEM) operating at 200kV equipped with a cold FEG and a Cs aberration probe corrector. All the analyses were performed using a probe convergent semi-angle of 29.4 mrad and a collection angle between 109 and 200 mrad allowing STEM-HAADF Z-contrast imaging.

Low temperature PL at 10K was performed using the 244 nm line of a frequency-doubled Ar laser with a spot diameter of 130 µm as the excitation source.

Spatially resolved micro(µ)-PL measurements were conducted at 65K using a continuous wave laser emitting at 355 nm, focused onto a $a_0 \approx 1.5$ µm-radius spot on the sample surface. The PL images were collected by a microscope objective that allows for 10 times magnification, and filtered through a vertical slit, which selects a 250 × 1 µm area on the sample surface. These spatial areas were analyzed by a spectrometer equipped with a 1200 lines/mm grating and a 2048 × 512 pixels CCD camera. The PL images that can be obtained with this set up are characterized by spatial and spectral resolution of ≈ 1 µm and ≈ 1 meV, respectively.

# 3   Sample growth and surface characterization

The surfaces of GaN (or AlGaN with a low Al composition typically below 0.2) grown by MBE using ammonia as the nitrogen source under N-rich conditions are generally constituted by hillocks (or mounds)[50] which are not related to a spiral growth but rather to the existence of a step-edge barrier for the diffusion of adatoms (Ehrlich-Schwoebel barrier).[51,52] This growth regime is unstable and induces a coarsening of the mounds and an increase of the surface roughness when the epitaxial layer thickness increases.[50] In order to improve the surface morphology, a step flow growth regime would be preferable and one solution to reach this regime would be to increase the growth temperature.[53] This is the purpose of the study described below.

## 3.1   Series as a function of temperature (samples A-D)

Figure 1 shows the AFM images of the surface of the first series of samples (A-D). The growth temperature has a strong impact on the surface morphology: the room mean square (rms) roughness of these 10×10 µm$^2$ AFM images is 1.2 nm, 0.6 nm, 2.3 nm, 6.4 nm for growth temperatures of 743°C, 773°C, 796°C, and 818°C, respectively. For temperatures below 800°C (Figure 1 a and b), the surface is composed of mounds and step meandering (the meanders or valleys are perpendicular to the steps). For temperatures larger than 800°C (Figure 1 c and d), the surface becomes pitted because of the strong desorption rate of N and NH$_x$ species[54] and an insufficient ammonia flow to stabilize the surface. The size and the density of the pits increases with the temperature. The formation of pitted GaN surfaces grown by MBE at relatively high temperature was already observed.[51,55] This issue can be solved by increasing the ammonia flow to compensate the N desorption. With such strategy, the growth of GaN by ammonia-MBE in a quasi-step-flow regime at 920°C was demonstrated showing smooth surfaces with a rms roughness of 1.1 nm for 5×5 µm$^2$ AFM image.[53]

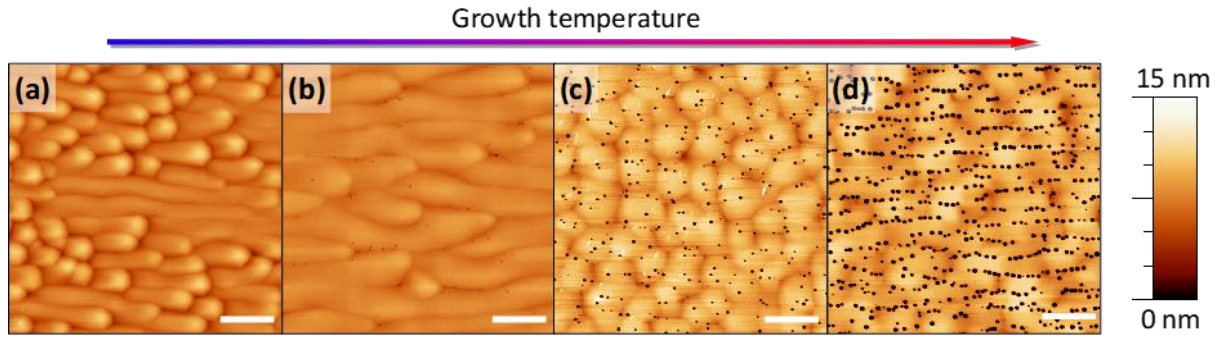

**Figure 1.** 10×10 µm² atomic force microscopy images of the AlGaN surface of samples A-D (a)-(d). The growth temperature increases from 743°C (left) to 818°C (right). The scale bar corresponds to 2 µm.

## 3.2  Improved process (samples A' and B')

In the second set of samples, the GaN substrates were systematically overgrown by a 1.1 µm-thick GaN layer by MOVPE. We observed that this kind of substrate preparation before MBE generally leads to smoother GaN/AlGaN surfaces in a more reproducible way than with the chemical surface preparation we used for the first set of samples. Sample A' is grown at a temperature close to the one of sample A and therefore the impact of the surface preparation can be evaluated. Sample B' is grown at 900°C with an ammonia flow which is 5 times larger than for the first set of samples.

Figure 2a shows a 5×5 µm² AFM image of the surface of a 600 nm-thick GaN layer grown with the same conditions as for sample B'. The surface is constituted by parallel steps indicating a step-flow epitaxial growth. The rms roughness is 0.1 nm which favorably compares with previous works about GaN homoepitaxy by ammonia-MBE.[53,55] The rms roughness of the top AlGaN surface for sample B' is 0.16 nm (Figure 2b). This larger rms roughness and the apparition of meanders (with a small peak-valley amplitude ~ 0.5 nm) is related to the lower surface diffusion of Al and also to the possible increase of the step-edge barrier for Al.

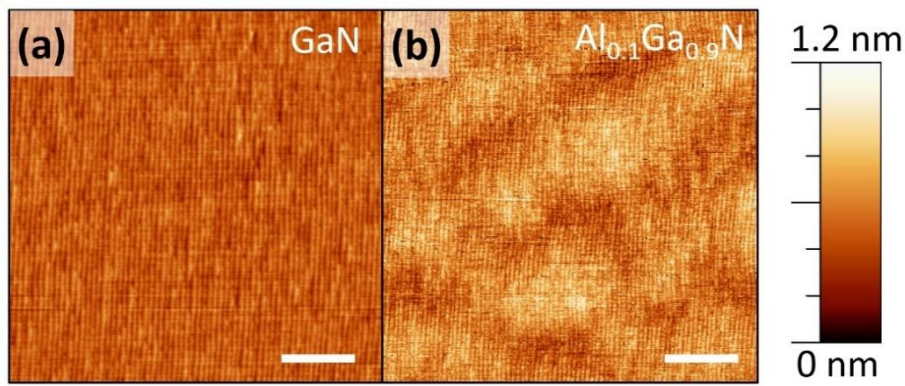

**Figure 2.** 5×5 µm² atomic force images of (a) a bare GaN surface grown at 900°C with an ammonia flow of 500 sccm and of (b) sample B' grown under the same conditions but with the GaN/AlGaN quantum well. The scale bar is 1 µm.

The surface properties of sample A' and B' are compared in Figure 3. The AFM image of sample A' (Figure 3a) is characteristic of a growth in the step meandering regime while that of sample B' (Figure 3b) approaches what is expected for a growth in the step flow regime. The rms roughness is 0.25 nm and 0.17 nm for samples A' and B', respectively. The corresponding height profiles are shown in Figure 3c. The peak-valley amplitude is about 1.5 nm for sample A' and 0.5 nm for sample B'. The height variation is quasi periodic with a period of 1.2 µm for sample A', and 2 µm for sample B'. We will see in section 4 that there are no clear correlations between surface roughness and exciton transport: despite larger peak-valley amplitude of the surface fluctuations on the micron scale, IX transport in Sample A' is more efficient than in Sample B'.

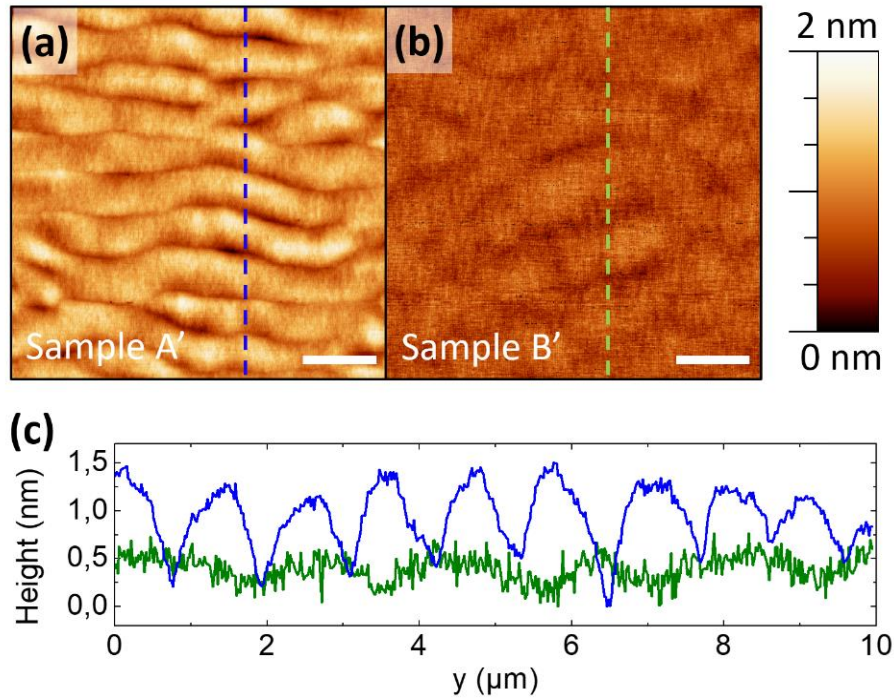

**Figure 3. Surface characteristics of samples A' and B'. 10×10 µm$^2$ atomic force microscopy images of sample A' (a) and B' (b). The scale bar is 2 µm. Height profile extracted along the line drawn in blue for sample A' and in green for sample B' (c).**

In order to determine whether the high temperature growth process can degrade the crystalline quality of the nitride layers, we have evaluated the dislocation density using panchromatic cathodoluminescence at room temperature.[56] No dark spots related to the presence of threading dislocations are visible on the CL panchromatic image of sample B' shown in Figure 4a, while a dark spot density of $1.7 \times 10^6$ cm$^{-2}$ is found for a sample with a similar structure grown on a free-standing GaN substrate (Figure 4b). According to the size of the image (100×100 µm$^2$), this represents a threading dislocation density lower than $10^4$ cm$^{-2}$, in line with the Ammono GaN substrate specifications.

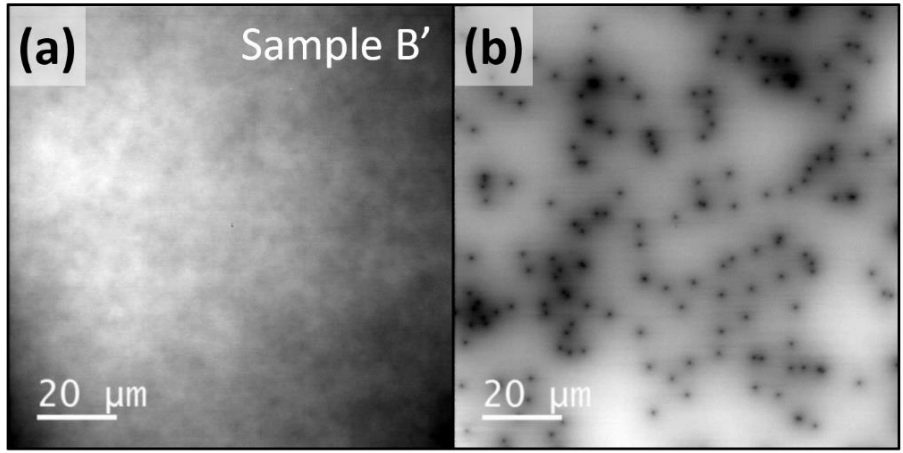

**Figure 4. Cathodoluminescence images in panchromatic mode at room temperature of sample B' (a) and for a sample with a similar structure grown on a free-standing GaN substrate (halide vapor phase epitaxy) (b).**

To obtain a better insight on the morphology and the crystalline quality of the epitaxial GaN/AlGaN QWs and on the presence of eventual defects at the interface we carried out high-resolution STEM-HAADF observation on cross-sections prepared using standard polishing TEM preparation method for samples A' and B'. Figure 5a and Figure 5c show low magnification STEM-HAADF images of the corresponding cross-section illustrating 0.2 µm long area representative of each sample. From these images we can infer that the GaN/AlGaN QWs present perfectly flat and continuous layers with no mixing at this scale.

A closer analysis of the GaN/AlGaN QWs for both samples at higher magnification represented in Figure 5b and Figure 5d allowed us to observe the defect free crystallinity of each layer individually, and the abrupt GaN/AlGaN interfaces.

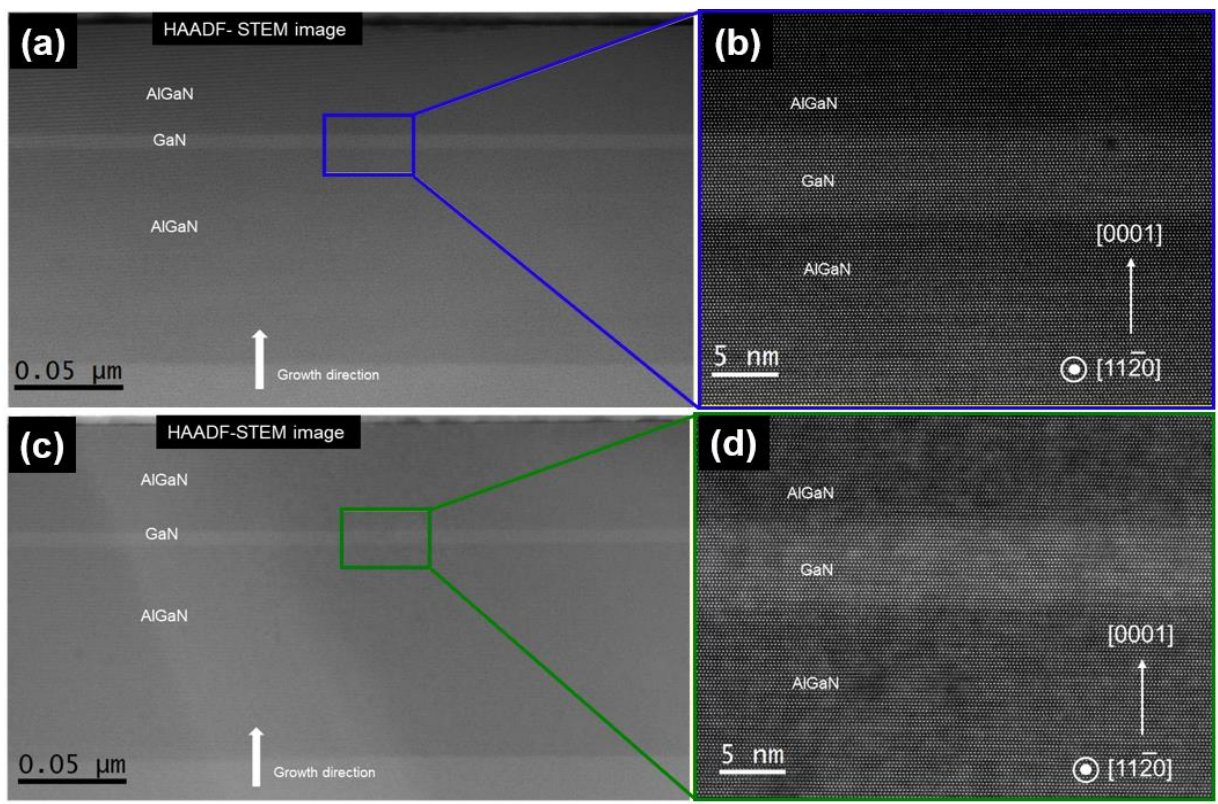

**Figure 5.** HAADF-STEM analysis of a cross-section of samples A' (a)(b) and B' (c)(d) prepared by standard polishing TEM preparation method; (a)(c) low magnification HAADF-STEM image of a thin area within the cross-section illustrating the morphology of the GaN/AlGaN QWs; (b)(d) HR-HAADF-STEM image representing a zoom on the GaN/AlGaN QWs interface marked by the rectangle area.

## 4 Photoluminescence properties

The PL properties were only evaluated for the samples A' and B' which possess the best surface morphologies compared to the first set of samples. Two different PL setups were used for this evaluation, see Section 2. The first one consists of a classic PL setup (macro-PL) at low temperature (10K) with a large diameter spot (130 µm) of the excitation laser. The results of these experiments can be easily compared with literature data. The second setup is more specific as it is a micro-PL setup at 65K equipped with a CCD camera allowing it to detect the diffusion of excitons within tens of micrometers range.

Note that sample areas studied at 10 K and 65 K are not exactly the same and correspond to different distances from the wafer center, so that the Al composition slightly differs in these two sets of experiments. From the measured low power IX emission energies, we estimate this difference to be less than 1%.

### 4.1 Photoluminescence at 10K with a laser spot diameter of 130 µm

Figure 6a shows the 10K PL spectra of samples A' and B' at an excitation power density of 0.7 W/cm$^2$. The PL spectra are dominated by the QW emission. The corresponding electronic state is an IX, the transition energy is situated close to 3.2 eV and accompanied by the longitudinal-optical phonon replica at 92.5 meV lower energy. PL peaks attributed to the GaN buffer layer and the AlGaN barrier layer are also visible at ~3.47 eV and 3.67 eV, respectively. The PL peak from the QW of sample A' is at 3.242 eV, while it is at 3.213 eV for sample B'. This difference is due to the larger Al composition of the AlGaN barrier in sample B' which induces a larger internal electric field and therefore a stronger quantum confined Stark effect.

The IX emission intensity in sample A' is 3.2 times larger than Sample B'. This could be due to a reduction of the exciton oscillator strength in sample B', as the quantum confined Stark effect is larger for this sample. However, calculations indicate (see section 4c) that the oscillator strength of the fundamental QW transition of sample A' is only 1.6 times larger than sample B'. Therefore, it cannot account for the PL intensity difference between two samples if we assume identical non-radiative recombination. Therefore, we conclude that the non-radiative recombination rate is more important in Sample B' than in Sample A'. This is not related to extended defects such as dislocations (see Figure 4a) and should be attributed to the presence of a larger density of non-radiative point defects in sample B'.

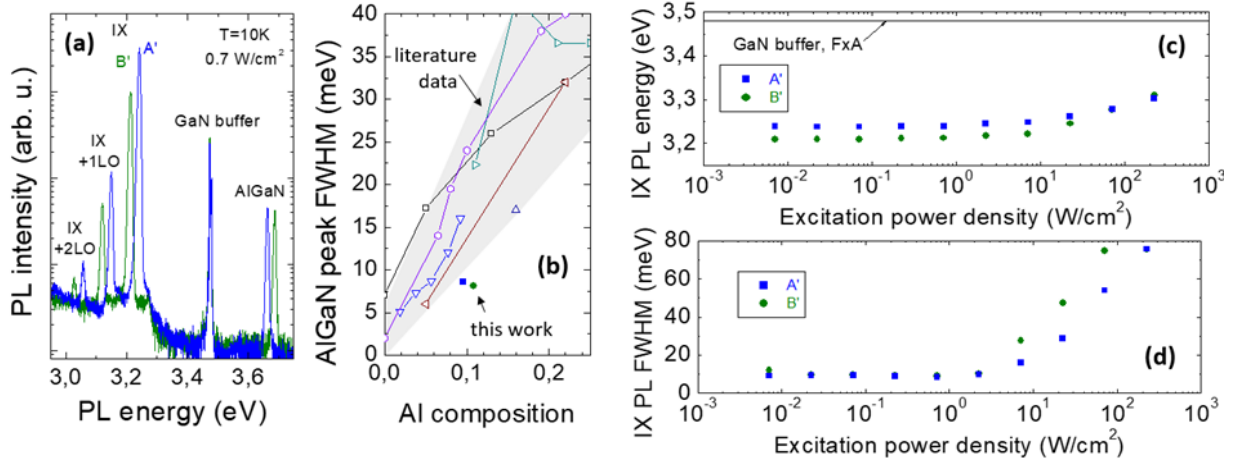

**Figure 6.** Low temperature photoluminescence (10K) using a 244 nm laser with a spot diameter of 130 μm as the excitation source for samples A' (growth temperature 780°C) and B' (growth temperature 900°C). (a) Photoluminescence spectra in log scale with an excitation power density of 0.7 W/cm². (b) Full width at half maximum of the AlGaN photoluminescence peak of samples A' and B' compared to literature data.[16,57,58,59, 60,61,62] Excitation power density dependence of the photoluminescence peak energy (c) and full width at half maximum (d).

The homogeneity of the AlGaN alloy can be evaluated by measuring the full width at half maximum (FWHM) of the corresponding PL peak. The FWHM is 8.6 and 8.1 meV for samples A' and B', respectively. These values are comparable with the best results published in the literature (Figure 6b).[16,57,58,59, 60,61,62]

Figure 6c and Figure 6d show the excitation power dependence of the PL peak energy and the FWHM of the IX emission. IX peak shifts towards high energies due to the progressive screening of the internal electric field by the photogenerated carriers whose density increases with the excitation power.[63]

The PL linewidth increases with the excitation power density due to the exciton-exciton scattering processes which becomes more probable when the exciton density increases. The PL broadening as a function of the power density is faster for sample B' compared to A' which can be explained by the larger carrier density for a same excitation power density. This issue will be addressed in greater detail in the last section. The minimum of the IX emission FWHM is 8.3 and 8.9 meV for samples A' and B', respectively. They are reached at an intermediate power density of 0.7 W/cm². At lower power densities, the IX emission linewidth slightly increases. We attribute this effect to the inhomogeneous broadening induced by IX localization. Measuring FWHM of the IX emission allows us to evaluate the homogeneity of the AlGaN alloy. We estimate that rms fluctuation of the Al fraction do not exceed 0.005.

Overall, the optical properties of these samples are very close, except the larger QW PL intensity of sample A'.

### 4.2  Micro-photoluminescence at 65K

Figure 7 shows PL spectra of sample A' (a-d) and B' (e-h) color-encoded in logarithmic scale. The spectra are taken at different excitation power densities, and the excitation spot is positioned at x = 0. The QW exciton emission and its longitudinal optical (LO) phonon replica (IX+1LO) can be readily identified. Indeed, their energies decrease with the distance from the excitation spot, reaching a constant value on the scale of several tens of micrometers, depending on excitation power. This behavior is consistent with previous observations on similar samples and is due to density-dependent screening of the built-in electric field: at highest power and at x=0, where maximum exciton density is reached, the highest emission energy is observed.

Dipole-dipole interaction between excitons pushes them away from the excitation spot, leading to the expansion of the exciton cloud. Thus, due to radial dilution via drift, diffusion, and recombination, IX emission energy decreases with increasing distance from the excitation spot. IX emission energy observed in the limit of large distances (x>30μm), where it does not change any more with the distance, can be interpreted as the "zero-density" exciton energy, $E_0$. The "zero-density" energy in sample A' is higher than in sample B', due to the larger Al composition of the AlGaN barrier in sample B' which induces a larger internal electric field $F$, see Table 2. This confirms the conclusions drawn from macro-PL experiments presented in the previous subsection.

Another important difference between the samples is the cloud expansion length. It is highlighted by arrows in Figure 7. This length is known to be strongly dependent on the excitation power,[35] and this is clearly observable in Figure 7 for both samples. One can also see that at any excitation power excitonic emission persists at larger distances in Sample A' than in Sample B'.

Finally, at ≈ 3.298 eV a weak but narrow emission line appears in almost the entire spatial area shown in Figure 7. It is identical in the two samples. We assigned this emission to the second LO replica of the free exciton in the GaN buffer layer at x=0, this light is guided in the sample plane and scattered by surface imperfections at x≠0.

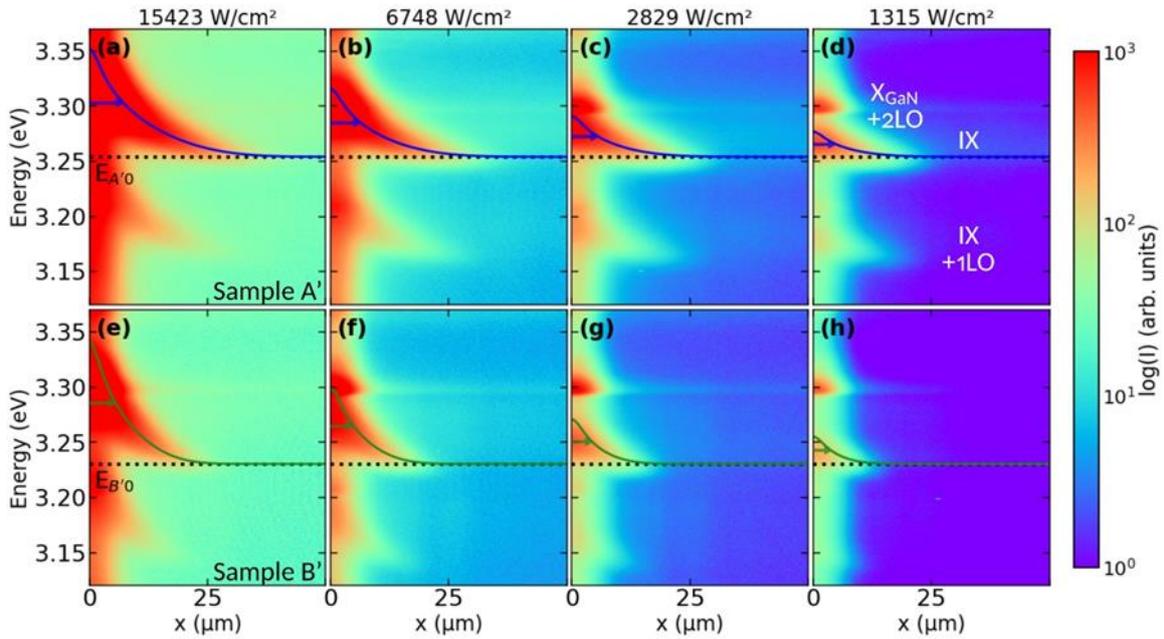

**Figure 7. Spatially resolved μPL spectra (color-encoded in log scale) for sample A' (a-d) and B' (e-h) at different excitation power densities (indicated on top of each column). Dotted lines indicate 'zero-density' energies, $E_0$. Solid lines are the results of the modeling by Eq. (4). Arrows indicate the extension of the IX cloud that is power dependent, but systematically larger for Sample A' then for Sample B'.**

### 4.3 Modeling of the exciton transport

In this subsection we present the modeling of the IX transport that allows us to identify and quantify the parameters that determine the observed characteristics of IX transport. The modeling procedure consists of three steps. Firstly, each spectrum corresponding to the distances larger than 5 µm is modeled by a phenomenological fitting function introduced in Ref. 37:

$$I = A \frac{\exp(\beta_l(E-E_c))}{1+\exp(\beta_h(E-E_c))} \quad (1)$$

This allows to determine IX peak energy $E_{IX}$ (slightly different from $E_c$), integrated intensity of the IX emission $I_{IX}$, and the linewidth, that we determine as full width at half maximum (FWHM) of the spectral line. $\beta_l(\beta_h)$ characterize the shape of lower (high) wing of the IX spectrum. Two spectra corresponding to the narrowest emission line for the two samples are shown in Figure 8(a), the shaded area underlines zero-phonon IX emission. Excitation power density and distances from the excitation spot are indicated in the figure caption.

Second, all the values of integrated intensity and FWHM measured at different excitation power densities and positions are reported as a function of the peak energy. The former dependence can be described within the model of built-in electric field that is screened by the increasing number of IXs. This leads to the increase of both energy (due to exciton-exciton interaction) and intensity (due to increasing overlap of electron and hole wave function within IX). Such behavior can be quantified by solving self-consistently Schrödinger and Poisson equations [we use NextNano software for numerical solution][64] for each sample. It appears that IX energy increases linearly with IX density, $n$:

$$E_{IX}(n) = E_0 + \phi_0 n, \quad (2)$$

where $\phi_0$ characterizes exciton-exciton interaction, and $E_0$ is the "zero-density" exciton energy. IX intensity increases exponentially with $n$:

$$I_{IX}(n) \propto n\Omega_0 \exp(n/\gamma), \quad (3)$$

where $\Omega_0 = |\int dz\, \psi_e(z)\psi_h(z)|^2$ is the squared electron-hole overlap integral at the zero-density excitonic transition and $\gamma$ characterizes the exponential increase of this overlap with increasing exciton density. Determination of $\Omega_0$, $\phi_0$ and $\gamma$ for various possible energies $E_0$ (or, equivalently, built-in electric fields) by solving self-consistently Schrödinger and Poisson equations allows us to fit Eq (2) to the dependence $I_{IX}(E_{IX})$ shown in Figure 8(b). From the best fit we deduced $E_0$, $\Omega_0$, $\phi_0$ and $\gamma$ for both samples. The result is shown by solid lines in Figure 8(b), and the parameters are given in Table 2.

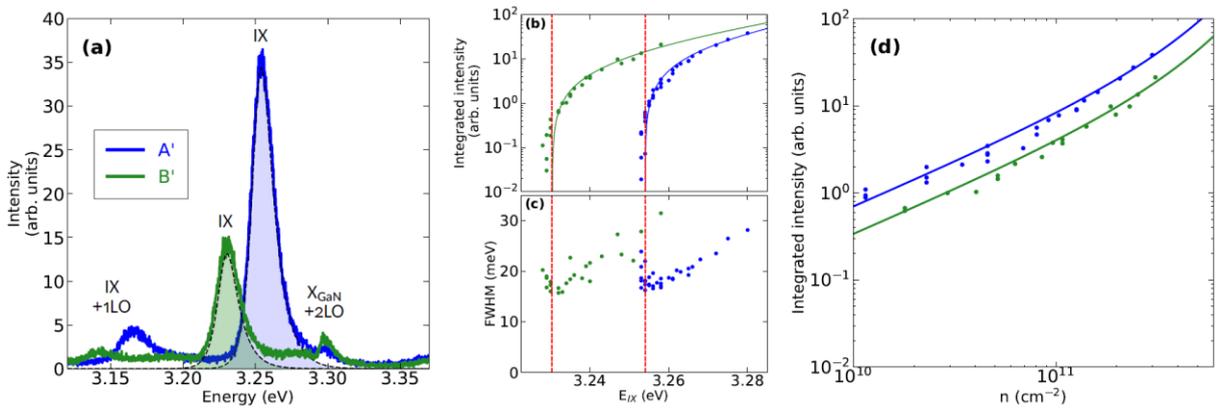

**Figure 8. (a) µPL spectra measured at x=17.0 µm (Sample A', blue) and x=10.8 µm (Sample B', green) at P= 740 W/cm. Spectrally integrated intensity (b) and linewidth (c) of the IX emission in sample A' (blue) and B' (green) as a function of the IX energy. Different points correspond to various positions in the plane of the QW and various excitation power densities (same data as in Figure 7). Solid lines are fitted to the data using Eq (3). Red dashed lines indicate $E_0$ for two samples. (d) Same data and modeling as in (b) but represented as a function of the IX density.**

The energies $E_0$ are indicated in Figure 8(b) by vertical dashed lines. One can see that $E_{0A'}>E_{0B'}$. This explains greater IX emission intensity in sample A', see Figure 8(d). At a given IX density the emission intensity ratio $I_{IXA'}/I_{IXB'} \sim 2$ is almost constant and is very close to the ratio $\Omega_{0A'}/\Omega_{0B'} \sim 1.6$. The fact that $I_{IXA'}/I_{IXB'}$ does not depend on the IX density despite exponential density dependence of the radiative rate corroborates the result of the macro-PL analysis, that pointed out the dominance of the non-radiative emission with respect to the radiative one. We will make more quantitative estimation of the corresponding times later on.

We should also comment on the origin of the excitonic emission measured at energies below $E_0$. We interpret this emission as being due to localized excitonic states, typical for nitride QWs,[22] but that can't be accounted for within the model presented above. A non-monotonous behavior of FWHM as a function of energy corroborates this interpretation, see Figure 8(c). Indeed, FWHM minimum is reached at energy $E_0$. At $E_{IX}<E_0$, FWHM increases due to localization-induced disorder (inhomogeneous broadening). Above $E_0$, FWHM increases due to enhanced exciton-exciton scattering.[65,66] We come back to this point in section 5 and compare the power-induced emission broadening measured in micro- and macro-PL experiments.

**Table 2. Parameters obtained from the self-consistent solution of Schrödinger and Poisson equations, from the modeling of $I_{IX}(E_{IX})$ dependence, and from the modeling of the IX transport.**

| Parameter | Definition | Units | Sample A' | Sample B' | Origin |
|---|---|---|---|---|---|
| $E_0$ | "Zero-density" energy | eV | 3.254 | 3.230 | Fit $I_{IX}(E_{IX})$ |
| L | Mean free path | nm | 12.5 | 8.5 | Fit $I_{IX}(x)$, $E_{IX}(x)$ |
| D | Diffusion constant | cm²/s | 4.1 | 2.8 | From L, $D=Lv_{th}$ |
| $\Omega_0$ | Squared overlap of the electron/hole wavefunctions | $\times 10^{-4}$ | 2.5 | 1.6 | Schrödinger-Poisson |
| $t_{rad}^0$ | Radiative recombination time at $n=0$ | μs | 0.42 | 0.65 | Fit $I_{IX}(x)$, $E_{IX}(x)$ and $t_{rad, B'}^0 = t_{rad, A'}^0 \times \Omega_{0, A'}/\Omega_{0,B'}$ |
| $t_{nr}$ | Non-radiative recombination time | μs | 0.07 | 0.03 | Fit $I_{IX}(x)$, $E_{IX}(x)$ |
| α | Number of IX per incident photon | | 0.064 | 0.064 | [67] |
| $\Phi_0$ | Mean-field IX interaction energy | meV $\times$ cm² $\times 10^{-11}$ | 8.7 | 8.9 | Schrödinger-Poisson |
| γ | Density dependence of the IX emission intensity | cm⁻² $\times 10^{11}$ | 5.14 | 5.23 | Schrödinger-Poisson |
| F | Built-in electric field | kV/cm | 658 | 699 | Schrödinger-Poisson |

The third stage of the modeling consists of the quantitative analysis of the spatial IX emission patterns and determination of the IX diffusion coefficients in the two samples. We describe spatial distribution of IX density by a steady-state solution of the drift-diffusion equation including pumping and relaxation terms.[36,37][68]

$$\frac{\partial n}{\partial t} = -\nabla \cdot (\vec{J}_{diff} + \vec{J}_{drift}) + G - Rn. \quad (4)$$

Here $\vec{J}_{diff} = -D\vec{\nabla} n$ and $\vec{J}_{drift} = -\mu n \vec{\nabla}(\varphi_0 n)$ are diffusion and drift currents, respectively. We assume that IX drift is due to dipole-dipole repulsion, and that diffusion coefficient, D, is determined by the average distance between fixed scattering centers (D=$Lv_{th}$, where $v_{th}$ is the thermal velocity of IXs). Diffusion coefficient is related to IX mobility by Einstein relation $\mu = D/k_B T$. Two other terms account for the exciton pumping at power density $P$, $G = \alpha \frac{P}{E_l} \exp(-x^2/a_0^2)$ and for relaxation at rate R. The later consists of two contributions, radiative and non-radiative, $R = R_{rad} + R_{nr}$. $E_l = 3.49$ eV is the pumping laser energy, $a_0 = 1.5$ µm is the laser spot radius and $\alpha$ is the number of IXs created per incident photon at x=0. From the known values of GaN absorption coefficient, $\alpha$ can be estimated as $\alpha = 0.065$.[67] The radiative recombination rate is given by $R_{rad} = \frac{1}{t^0_{rad}} \exp(n/\gamma)$, where "zero-density" radiative time $t^0_{rad}$ is inversely proportional to $\Omega_0$, that we have determined previously. This imposes the relation between the radiative lifetimes in the two samples: $t^0_{rad, B'} = t^0_{rad, A'} \times \Omega_{0, A'}/\Omega_{0, B'}$. The non-radiative rate $R_{nr} = \frac{1}{t_{nr}}$ is density-independent and considered as a fitting parameter. The relation between IX density and energy is given by Eq. (2), and between density and emission intensity by $I_{IX} = n R_{rad}$. More details on the IX transport modeling by drift-diffusion equation can be found in Refs. 35,36. Below we present the steady-state solutions of Eq. (4) that we obtained and compare them with the experimental results.

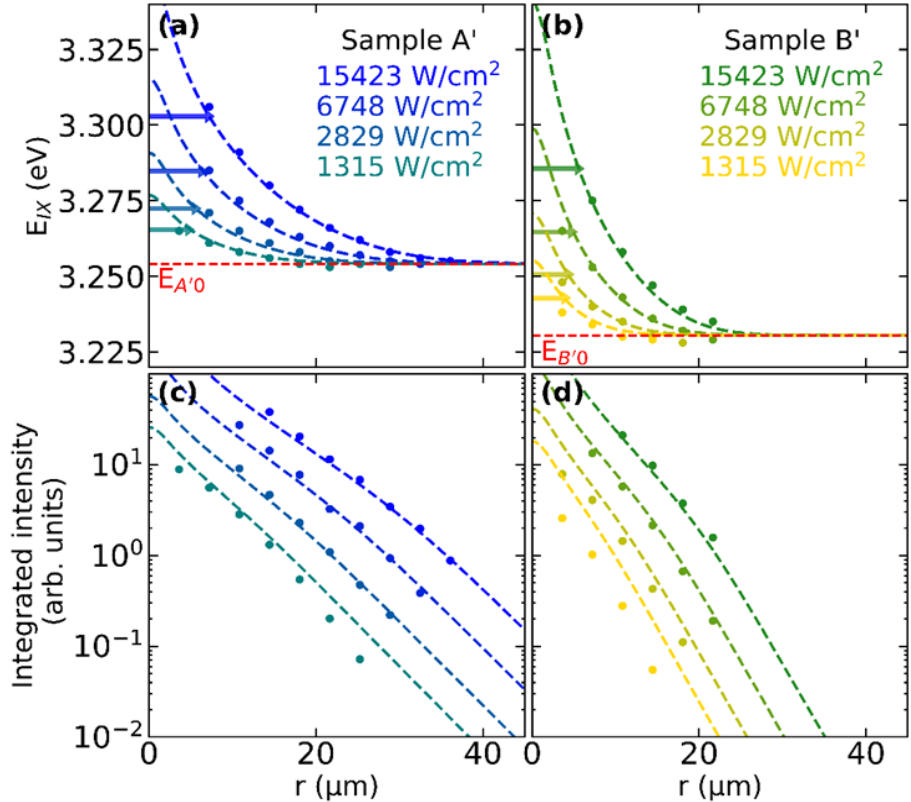

**Figure 9.** Exciton energy (a, b) and spectrally integrated intensity (c, d) as a function of the position in the QW plane (excitation spot is situated at x = 0) in samples A' (a, c) and B' (b, d). Dashed lines are obtained from steady-state solutions of Eq. (4) as $E_{IX} = E_0 + n\phi_0$, $I_{IX} = nR_{rad}$. Arrows indicate the extension of the IX cloud that is power dependent, but systematically larger for Sample A' then for Sample B'.

Figure 9 shows IX emission energy (a, b) and intensity (c, d) as a function of the distance from the excitation spot measured in Sample A' (a, c) and Sample B' (b, d). These data (solid circles) are extracted from the spectral maps like those shown in Figure 7. Such presentation allows us to compare in the same figure IX energy and intensity profiles at various power densities. Dashed lines represent the steady state solutions of Eq. (3). The values of the fitting parameters that we obtained ($t^0_{rad, A'}$, $t_{nr}$, and L) as well as the deduced values ($t^0_{rad, B'}$ and D) are given in

Table 2. Note that, we are able to determine independently the recombination times and the diffusion constants because we fit the model to both energy and intensity data.

The main outcome of this analysis is that the more efficient IX transport that we observe in Sample A', is due to a combination of two factors: smaller non-radiative losses, which were already evidenced by macro-PL experiments, but also larger diffusion coefficient. It seems reasonable to suppose that Sample A' presents a smaller density of point defects that leads to both longer IX mean free path L, and smaller non-radiative recombination rate.

## 5   IX emission broadening in PL and µPL

The linewidth of optical emission is an important parameter characterizing QW quality. This section is devoted to comparison of the IX emission broadening that we measure in two types of experiments in Samples A' and B'. To make this comparison meaningful, we represent the line broadening not as a function of incident power, but as a function of emission energy, which, in first approximation, is proportional to the IX density.[63] Figure 10 shows the FWHM as a function of the corresponding IX emission peak energy measured in macro-PL and µPL experiments, at 10K and 65K, respectively.

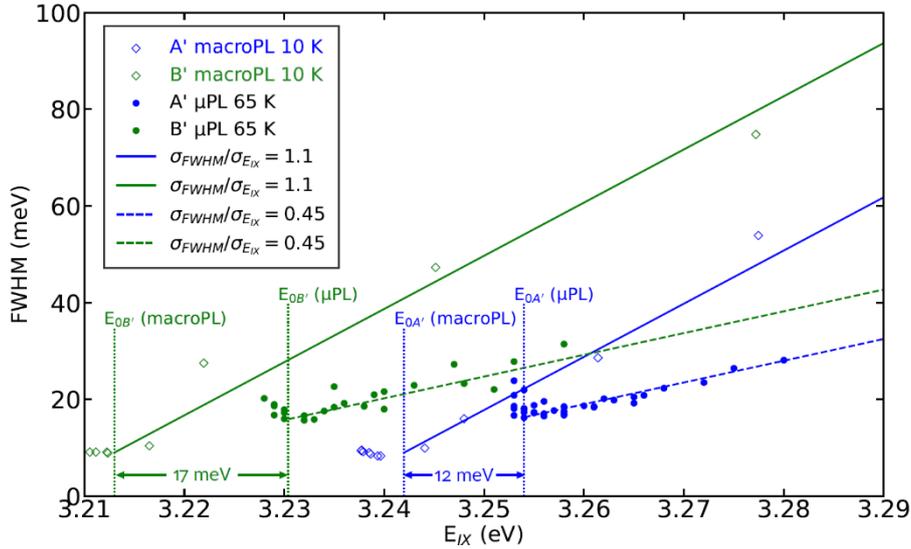

Figure 10. FWHM of the IX emission line as a function of its peak energy extracted from µPL at 65K (solid circles) and macro-PL at 10K (open diamonds). Lines are linear fits of the line broadening above $E_0$. Vertical dotted lines indicate the values of zero-density energy, $E_0$, for each set of data. Differences between the values of $E_0$ measured in the center (macroPL) and in the periphery (µPL) of the wafer are highlighted by arrows.

As discussed in the previous section, the minimum FWHM values are reached not at the lowest IX densities, but at "zero-density" energy $E_0$, where localized states are occupied but there are only a few free IX in the QW. This is observed in both macro-PL and µPL. These minimum values are very close in two samples: ~9 meV at 10K measured in macro-PL and 17 meV at 65K measured in µPL experiments. This indicates that surface roughness amplitude, which differs strongly in these samples (1.5 nm in Sample A' and 0.5 nm in Sample B', see Figure 3, is not the dominant mechanism of the IX inhomogeneous broadening. Most probably, it is determined by the fluctuations of the Al content in the barriers, which are similar in our two samples. We estimate that fluctuations as small as ±0.2% could result in inhomogeneous broadening of 9 meV.

The ratio between minimum values of FWHM measured at 10K and 65K is consistent with temperature-dependent inhomogeneous broadening if we suppose that for free excitons it is proportional to the QW area, A, explored during IX lifetime $A \propto D \propto \sqrt{T}$. In this case we expect that FWHM$\cong$17 meV at 65K would be reduced to FWHM$\cong$7 meV at 10K, close to FWHM$\cong$ 9 meV deduced from macro-PL experiments.

The values of $E_0$ slightly differ in two types of experiments (by 12 meV in Sample A' and by 17 meV in Sample B', see arrows in Figure 10). This is because the temperature is different and the sample areas studied by macro-PL and µPL correspond to different areas of the wafer, which results in a small difference in Al composition in the barriers (less than 1%).

At low IX densities, corresponding to $E_{IX}< E_0$, IX localization leads to the increase of the linewidth. This inhomogeneous contribution to the broadening is better resolved in µPL but also visible in macro-PL data. At high IX densities, which correspond to $E_{IX}>E_0$, the FWHM increases linearly with the energy. As mentioned in Section 4.3, one can expect such increase to be due to increased exciton-exciton scattering.[65] At very high densities, close to Mott transition and above (for our samples we estimate the corresponding blueshifts as $E_{IX}$- $E_0>$30 mV), which were only reached in macro-PL experiments, the line can be additionally broadened by a step-like electron and hole joint density of states.[37,69] Here we observe that in µPL the slope σ=(FWHM($E_{IX}$)- FWHM($E_0$))/($E_{IX}$-$E_0$)$\cong$0.45 is the same for the two samples, and quite close to the results that we reported previously.[66] In macro-PL, however, this slope is more than twice larger, σ$\cong$1.1. This can be easily understood. Indeed, IX energies strongly depend on their densities. Therefore, an inhomogeneous IX density profile created by a broad Gaussian-like spot on the surface will inevitably lead to a broad distribution of emission energies at different positions. In the absence of spatial resolution, such distribution appears as an additional inhomogeneous broadening, which increases with IX density. This effect must be accounted for when analyzing QW quality.

# 6  Conclusions

We have grown a set of GaN/AlGaN QWs to optimize IX transport in the QW plane. Surface and interface roughness, cathodoluminescence, PL, and spatially resolved µPL have been scrutinized in these samples.

By analyzing the surface morphology, we evaluated the effects of the substrate growth temperature on the growth mode (step flow versus mound formation) and dislocation densities. Further optimization has been reached by additional MOVPE growth of 1.1 µm-thick GaN layer on the substrate prior to MBE growth of the QW heterostructure. As a last step, we have comparatively studied optical emission of the two best samples obtained within the same growth procedure but keeping the substrate at different temperatures, 780°C (Sample A') and 900°C (Sample B').

Summarizing the results of optical experiments, we conclude that the growth protocol used for Sample A' is the most advantageous for studies of IX. In this sample macroPL experiments point out lower non-radiative losses, while µPL experiments confirm this observation and demonstrate IX transport over larger distances. Thus, IXs have greater diffusion constant and mobility in Sample A' than in Sample B'. The corresponding mean free path of IXs are estimated as ~13 and ~9 nm in Samples A' and B', respectively. We tentatively interpret these values as an average distance between scattering centers which are also responsible for the non-radiative losses. These centers are not related to extended defects, such as dislocations since the latter are characterized by much lower densities. The corresponding diffusion constants at T=65K are D~4 cm$^2$/s and D~3 cm$^2$/s for Samples A' and B', respectively. These values are close to those found in GaAs/AlGaAs QWs of the similar thickness by Vörös et al.,[65] but about an order of magnitude smaller than those reported in ref. [70]. Importantly, the mechanisms that limit exciton diffusion are different in GaN and GaAs-based heterostructures. While in GaAs/AlGaAs QWs the IX transport is mainly limited by the QW interface roughness, in GaN/AlGaN QWs the non-radiative defects seem to be a major limiting factor.

In contrast with IX diffusion constants, the linewidth of the IX emission is essentially the same in Sample A' and B'. We speculate that the latter is determined by Al content fluctuations in the barriers, which are much less sensitive to the growth temperature than the density of non-radiative defects.

Overall, the studied samples are promising for studies of many-body physics of dipolar bosons. They could be patterned with metallic electrodes in order to make traps able to confine spatially IX as this was demonstrated in previous work.[36,66] We expect that thanks to the improved epitaxial quality of the samples and the reduced excitonic dipole length, quantum collective effects such as Bose-Einstein condensation could be demonstrated at temperatures above 1 K.

Finally, several other avenues can still be explored to improve samples quality. In particular, the AlGaN alloy fluctuations that are the main source of the broadening of the IX emission, could be decreased. The straightforward way to do that would be to decrease the Al composition in the AlGaN barrier layer. Indeed, an AlGaN PL FWHM down to 5 meV was demonstrated for an Al composition as low as 0.05 (Figure 6b).[16] However, the limit of this approach is that the exciton confinement in the QW may decrease, which can induce the exciton escape from the QW especially for large carrier densities and for non-cryogenic temperatures. Other approaches could be considered, such as the optimization of the substrate offcut[71,72] or the use of short period AlN/GaN superlattices (AlGaN digital alloy) to further minimize the impact of AlGaN alloy fluctuations.[73,74,75]

## Data availability

The data that support the findings of this study are available from the corresponding author upon reasonable request.

## Acknowledgments

This work was supported by the project IXTASE of the French National Research Agency (ANR-20-CE30-0032) and by the Occitanie region through the Quantum Technology Challenge grant. This work also benefited from the aid from the French government, managed by the National Research Agency under the project "Investissements d'Avenir" UCA JEDI (ANR-15-IDEX-01).